\documentclass[preprint,showpacs,preprintnumbers,amsmath,amssymb]{revtex4-1}         %
\usepackage{graphicx,epsfig}% Include figure files
\usepackage{dcolumn}% Align table columns on decimal point
\usepackage{bm}% bold math
\newcommand{\be}{\begin{equation}}
\newcommand{\ee}{\end{equation}}
\newcommand{\bse}{\begin{subequations}}
\newcommand{\ese}{\end{subequations}}
\newcommand{\bary}{\begin{eqnarray}}
\newcommand{\eary}{\end{eqnarray}}
\newcommand{\bwt}{\begin{widetext}}
\newcommand{\ewt}{\end{widetext}}
\begin{document}

\title{Detection of ultra high energy neutrinos by IceCube: Sterile neutrino scenario}
\author{Subhash Rajpoot$^{*}$, Sarira Sahu$^{**}$, and Hsi Ching Wang$^{*}$
}
\affiliation{
$^{*}$Department of Physics \& Astronomy, California State
University,
1250 Bellflower Boulevard, Long Beach, CA, 90840, USA\\
$^{**}$Instituto de Ciencias Nucleares, Universidad Nacional Aut\'onoma de M\'exico,
Circuito Exterior, C.U., A. Postal 70-543, 04510 Mexico DF, Mexico}
\email{Subhash.Rajpoot@csulb.edu}
\email{$^{**}$sarira@nucleares.unam.mx}
\email{$^*$Hsi-ching.Wang@cgu.edu}

%\date{\today} % It is always \today, today,
             %  but any date may be explicitly specified

\begin{abstract}

The short-baseline neutrino oscillation experiments, the excess of
radiation from the measurement of the cosmic microwave background
radiation, the necessity of the nonbaryonic dark matter candidate and
the depletion of the neutrino flux in IceCube all seem to hint at new
physics beyond the standard model.
An economical way to address these issues is to invoke the existence
of \emph{sterile neutrinos}.
We present simple extensions of the standard model with additional three sterile
neutrinos and discuss the corresponding PMNS like neutrino flavor
mixing matrix. The noteworthy features of the sterile neutrino scenario advocated
here is that the lightest one is almost degenerate with one of the active neutrinos,
the second sterile has mass of order eV  and the heaviest one is
 in the keV range. In the present scenario,  the short baseline anomaly
 is explained through $\Delta m^2\sim 1\, {\rm eV^2}$, the depletion
 of muon neutrino flux in IceCube is explained through $\Delta m^2\sim
 4.0\times 10^{-16}\, {\rm eV^2}$ and the dark matter problem is
 addressed through $\Delta m^2\sim 1\, {\rm keV^2}$.
Our proposed mixing matrix is also compatible with
the observed neutrino oscillation data. We show that the high energy
muon  and the tau neutrino fluxes from Gamma Ray Bursts can be
depleted in IceCube by as much as 38\% and 43\% respectively.
 These substantial depletion in  both muon and tau neutrino fluxes is
 due to their  small but sizable mixing with the sterile neutrinos.

\end{abstract}

\pacs{14.60.Pq; 14.60.St; 95.85.Ry} % PACS, the Physics and Astronomy
                       % Classification Scheme.
%\keywords{Suggested keywords}%Use showkeys class option if keyword
                              %display desired
\maketitle

\section{Introduction}

In the standard picture of neutrino oscillations, the three active
neutrino states  are
linear superpositions of three mass eigenstates. The
oscillation experiments with solar, atmospheric, reactor and
accelerator neutrinos can be explained through the mass squared
differences \cite{Abe:2010hy,Adamson:2011ig}
\bary
\Delta m^2_{\rm sol}\equiv\Delta m^2_{21}\simeq 7.6 \times 10^{-5} eV^2,\nonumber\\
\Delta m^2_{\rm atm}\equiv\Delta m^2_{31}\simeq 2.45 \times 10^{-3}
eV^2,
\label{dm2}
\eary
with $\Delta m^2_{ij}=m^2_i-m^2_j$.
The three mixing angles in this scheme have also been measured.  The solar \cite{Ahmad:2002jz}
and KamLand\cite{Abe:2008aa}  data give $\sin^2{\theta_{12}}\simeq 0.3$, the
atmospheric\cite{Fukuda:1998mi}  and MINOS\cite{Michael:2006rx}  data give $\sin^2{\theta_{23}}\simeq
0.5$. Also recently the Double-CHOOZ\cite{Abe:2011fz},
RENO\cite{Ahn:2012nd}  and Daya-Bay\cite{An:2012eh}
experiments measure the third mixing angle $\sin^2{\theta_{13}}\simeq 0.1$.
However, the completeness of the three-neutrino mixing paradigm is
in question due to several anomalies observed in the appearance
and disappearance of neutrinos in data pertaining to  short
baseline (SBL) experiments; Liquid Scintillator neutrino Detector (LSND)\cite{Aguilar:2001ty},
Mini-Booster Neutrino Experiment (MiniBooNE)\cite{AguilarArevalo:2010wv} and the reactor
anomaly\cite{Mention:2011rk} (henceforth all combined and referred to as SBL anomaly).
The SBL anomaly can not be accommodated with just three
active neutrinos, thus suggesting the possible existence of
one or more eV-scale sterile neutrinos to explain these results\cite{ALEPH:2005ab}.

Although the existence of dark matter (DM) in the Universe is confirmed
beyond doubt, its nature is still an outstanding puzzle both in
particle physics and cosmology. To be consistent with the
observations, the DM candidate should be a very weakly interacting,
electrically neutral particle. Sterile neutrinos with a mass ${\cal
  O}(1)$ keV and lifetimes much longer than the age of the Universe
are very good candidates for warm dark matter
(WDM)\cite{Kusenko:2009up,Boyarsky:2009ix}.  These sterile neutrinos
could be produced in the early Universe and their mass is generated by a
Majorana mass term which is not bound to the electroweak scale. Apart
from explaining the DM problem sterile neutrinos may also explain the large pulsar
kick velocity\cite{Fuller:2003gy}, and may also suppress the formation
of dwarf-galaxies and other small-scale structures.

Gamma-ray bursts (GRBs) and active galactic nuclei (AGN) are believed to be the prime candidates for
the production of ultra-high energy cosmic rays (UHECRs) and ultra-high
energy neutrinos are their by-products\cite{Halzen:2002pg,Becker:2007sv,Murase:2008mr,Murase:2005hy}. IceCube, the ${\rm km}^3$-scale
neutrino telescope  constructed in the South Pole is meant to detect
these cosmological neutrinos\cite{Halzen:2010yj}. The IceCube collaboration recently
published their analysis of data taken during the construction phase
using the 40- and 59-string configurations of the detector. The
combined analysis of the data does not show any neutrino signal
correlated with the observed GRBs during the data taking period\cite{Abbasi:2012zw,Abbasi:2011qc}. From
this analysis, IceCube has set an upper bound on the neutrino flux
from GRBs which is at least a factor of
3.7 below the Waxman-Bahcall (WB) predication\cite{Waxman:1997ti}. 
This depletion in the neutrino flux gave rise to many possible
explanations\cite{Chikashige:1980qk,Gelmini:1980re,Beacom:2002vi}.

From the astrophysics point of view,
it has been pointed out recently that for the normalization of neutrino flux,
IceCube ignored the effects of energy dependence of charged pion
production and secondary pion/muon cooling in the GRB fireball which caused an
overestimation of neutrino flux by a factor of 5 for typical GRB
parameters\cite{Li:2011ah}. Furthermore, by taking into account many other effects
(pion and kaon production
models, magnetic field effect and neutrino flavor mixing)  and doing a
full numerical calculation it is showed
that neutrino flux reduces by about one order of
magnitude\cite{Hummer:2011ms}. With the revised neutrino flux
calculation, reduction
in flux is also obtained by analyzing neutrino
flux from 215 GRBs during the period of 40- and 59-string
configuration of the IceCube\cite{He:2012tq}.  There are also
alternative astrophysical models \cite{Gao:2012ay,Baerwald:2013pu,Zhang:2012qy} which predict a lower neutrino flux
compared to the WB models. So the claim by IceCube
may not be that serious, but the WB models can
be challenged in future as the observations put stringent limits on
the muon neutrino flux.

To address this issue from the particle physics point of view, the
existence of pseudo-Dirac neutrinos\cite{Wolfenstein:1981kw,Petcov:1982ya,Bilenky:1983wt,Kobayashi:2000md,Cahill:1999yk} is postulated. In this scenario
the neutrino of
each generation is composed of an almost  maximally mixed active-sterile neutrino combination,
separated by a tiny mass difference so that the active-sterile
oscillations are possible without affecting the short baseline
oscillation results\cite{Beacom:2003eu,Esmaili:2012ac}.
In a recent paper it has been postulated that apart from the above
explanation, neutrino decay can also be a viable explanation for the
suppression of muon neutrino flux\cite{Pakvasa:2012db}. 
Yet another very recent paper discusses about
the suppression of muon neutrino flux in IceCube by assuming that all
the neutrinos are pesudo-Dirac in nature and there is a mirror world
replicating the interaction in the observed world and also connected
to the later gravitationally. In this scenario each active neutrino is
associated with three sterile neutrinos with a very tiny splitting and
oscillation from active to sterile can be responsible for the
suppression of muon neutrino flux\cite{Joshipura:2013yba}.
So, if at all sterile neutrinos exist, and one/some of them are closely
degenerate in mass with the active neutrinos, and also mix, they may
easily evade detection in oscillation experiments. However,
due to the very long baseline involved in the oscillation process
sterile neutrino, in principle, can have measurable effects on the
high energy neutrino flux. 
Also the possibility of  sterile neutrino was
looked for in the atmospheric neutrino data collected by AMANDA and
partially deployed IceCube\cite{Esmaili:2012nz}.

These postulated sterile neutrinos neither participate in the weak
interaction nor contribute to the invisible width of
the Z boson\cite{ALEPH:2005ab}.  Also there is no known fundamental
symmetry in nature forcing the existence a fixed
number of sterile neutrino species. Cosmological probes such as  bounds on the relativistic energy density of
the universe in terms of the effective number of light
neutrinos\cite{Komatsu:2010fb} have been
extensively used to set bounds on the number of light neutrinos in
general and the number of sterile neutrinos in particular.

%Interestingly,  all the above anomalies  may be resolved or ameliorated
%by introducing additional massive sterile neutrinos. In this context
%Many phenomenological models with one or more  sterile
%neutrinos have been postulated to address these
%problems\cite{Giunti:2011gz,Archidiacono:2012ri,Donini:2012tt,Keranen:2003xd,Biermann:2006bu,Conrad:2012qt,Dev:2012bd,Pakvasa:2012db}.

In this work,  we extend the Standard Model  to include three additional sterile
neutrinos (3+3). All neutrinos in the model, active and sterile,  have
nonzero  masses and mix. 
The flavor mixing among the neutral leptons  gives rise to a $6\times 6$
matrix, analogous to the PMNS scheme for the active neutrinos.  We
will show that the generalized $6\times 6$ 
 matrix is compatible with the observed active neutrino oscillation
data. Although our main focus is to explain the depletion of
muon neutrino flux in IceCube, our model also encompasses solutions to
the SBL anomaly and the dark matter problem.

%where the lightest one is
%almost degenerate with one of the active one, the intermediate one has
%mass ${\cal O} (1)$ eV and the heaviest one has mass ${\cal O} (1)$
%keV. Also we give the (
%This extension of SM is natural from the left-right symmetric model.
%%%%%%%%%%%%%%%%%%%%%%%%%%
%%%%%%%%%%%%%%%%%%%%%%%%%%
\section{The (3+3) model}
\label{model33}
%%%%%%%%%%%%%%%%%%%%%%%%%
%%%%%%%%%%%%%%%%%%%%%%%%%%
%%%%%%%%%%%%%%%%%%%%%%%%%%
We assume that apart from three active neutrinos
there are  three additional sterile neutrinos. The flavor states of
the active neutrinos are defined as $\mid \nu_{e} \rangle$,
$\mid \nu_{\mu} \rangle$ and $\mid \nu_{\tau} \rangle$. The corresponding
mass eigenstates are  $\mid \nu_{1} \rangle$,
$\mid \nu_{2} \rangle$ and $\mid \nu_{3} \rangle$ with masses $m_1$,
$m_2$ and $m_3$ respectively. The flavor states of
the sterile neutrinos are defined as $\mid \nu_{a} \rangle$,
$\mid \nu_{b} \rangle$ and $\mid \nu_{c} \rangle$. The corresponding
mass eigenstates are  $\mid \nu_{4} \rangle$,
$\mid \nu_{5} \rangle$ and $\mid \nu_{6} \rangle$ with masses $m_4$,
$m_5$ and $m_6$ respectively.

%%%%%%%%%%%%%%%%%%%%%So the sum goes from 1to 6.%%%%%%%%%%%%%%%%%
In the standard treatment of neutrino oscillation in vacuum, the flavor  and the
mass eigenstates are defined as $\nu_{\alpha}$ and $\nu_i$
respectively. The flavor states are superpositions of mass eigenstates with non-zero
mass square difference and are given as
\be
\mid \nu_{\alpha} \rangle = \sum^6_{i=1} U^*_{\alpha i} \mid \nu_i\rangle.
\label{nustate}
\ee
The mixing matrix $U$ is the extended
Pontecorvo-Maki-Nakagawa-Saki (PMNS) matrix.
The three lowest states $\mid \nu_{1} \rangle$,
$\mid \nu_{2} \rangle$ and $\mid \nu_{3} \rangle$ with their
respective masses $m_1$, $m_2$ and $m_3$ account for solar and atmospheric neutrino
oscillations. We assume that the  sterile neutrinos are
Majorana singlets. They could either be left handed or right handed.
Below we shall describe two models with Majorana steriles. The first model in which $\mid \nu_{a} \rangle$,
$\mid \nu_{b} \rangle$ and $\mid \nu_{c} \rangle$ are left handed will be referred to as model A while the second model in which $\mid \nu_{a} \rangle$,
$\mid \nu_{b} \rangle$ and $\mid \nu_{c} \rangle$ are right handed
will be referred to as model B.

We present a standard model extension in which the seesaw
mechanism is invoked to generate the required spectrum of  light
sterile neutrinos. The standard model
with three generations of quarks and leptons is extended in  the
leptonic sector to include three right handed neutrinos and three
vectorlike neutral leptons \cite{ArkaniHamed:2005yv,Mahbubani:2005pt,D'Eramo:2007ga,Enberg:2007rp}. In total, our model has twelve neutral
leptons, three left handed active neutrinos $ \nu_L $ $\equiv$ ($\mid
\nu_{eL} \rangle$, $\mid \nu_{\mu L} \rangle$, $\mid \nu_{\tau L}
\rangle$), their counter parts, the three right handed inert neutrinos
$ \nu_R$ $\equiv$ ($\mid \nu_{eR} \rangle$, $\mid \nu_{\mu R}
\rangle$, $\mid \nu_{\tau R} \rangle$), additional
three left handed and three right handed neutrals
$N_L$ $\equiv$ $(N_1, N_2, N_3)_L$, $N_R$ $\equiv$ $(N_1, N_2,
N_3)_R$.
The interaction Lagrangian relevant for neutrino masses and mixings is symbolically given by
\bary
-{\cal L}_Y &=&  \bar L_i Y\Phi\, \nu_{R}\: +\ \bar L_i Y^{'} \Phi\, N_{R}\:
 +\ \frac{1}{2}\bar{\nu}^C_{R} (M)_{}\nu_{R}\nonumber\\
&&
+\:\bar{N}_{ L}\, (\Delta^{'})_{} \nu_{R}
\: +\: \frac{1}{2}\bar{N}_{L}(M_{LL})_{} N^C_{L}\:
\: +\: \bar{N}_{R}(m_{RR})_{} {\nu}^C_{R}\: ~ \nonumber\\
&&
+\:\bar{N}_{L}\, (\Delta^{''}_{} )N_{R}
\: +\: \frac{1}{2}\bar{N}_{ R}
(M_{RR})_{} N^C_{R}\: +\ {\rm H.C.}~,
\eary
which gives rise to the following $12\times 12$ neutrino mass matrix
in the basis $\{\nu_{L},\nu_{R}^C,N_{L} ,N_{R}^C \}$:
\bary
{\cal M}_\nu\ =\ \left(\begin{array}{cccc}
{\bf 0} & \Delta & {\bf 0} & {\Delta^{'}} \\
\Delta^{\sf T} & M &  \Delta^{''\sf T} & m_{RR}^{\sf T} \\
{\bf 0} & \Delta^{''} & M_{LL} & {\Delta}_{LR} \\
{\Delta^{'{\sf T}}} & {m_{RR}} & {\Delta}_{LR}^{\sf T} & {M_{RR}}
\end{array}\right)\; .
\label{eq:inverse1}
\eary
All entries in ${\cal M}_{\nu}$ are $3 \times 3$ matrices.  $Y$ and $Y^{'}$ represent
Yukawa couplings,   $\Delta=Y \langle {\Phi}\rangle$ is  the Dirac
mass matrix  for the  active
neutrinos, $\Delta^{'}=Y^{'} \langle {\Phi}\rangle$ is  the Dirac
mass matrix  for the  active neutrinos and $N_R$.
$M$   is  the   $(B-L)$-breaking   Majorana  mass   matrix  of   the
 right handed neutrinos, $\Delta^{''}$ is  the Dirac  mass matrix  for
 $\nu_R$ and $N_L$.  $\Delta_{LR}$ is  the Dirac  mass matrix  for
 $N_L$and $N_R$.
 The remaining terms in ${\cal M}_{\nu}$  are all
 Majorana mass matrices. The model has enough parameters to give
 representative values
for masses and flavor mixings  to address the short-baseline neutrino oscillation experiments, the excess of
radiation from the measurement of the cosmic microwave background
radiation, the need for nonbaryonic dark matter   and
the depletion of the neutrino flux in IceCube. \\
%%%%%%%%%%%%%%%%%%%%%%%%%%%%%%%%%%%%%%%%%%%%
%%                   MODEL A              %%
%%%%%%%%%%%%%%%%%%%%%%%%%%%%%%%%%%%%%%%%%%%%
\noindent \textbf{Model A:} In this model the light mass eigenstates
are the active states $ \nu_L $ $\equiv$ ($\mid \nu_{eL} \rangle$,
$\mid \nu_{\mu L} \rangle$, $\mid \nu_{\tau L} \rangle$) and the
sterile neutrals $N_L$ $\equiv$ $(N_1, N_2, N_3)_L$.  The lightness of
the states is achieved by invoking the seesaw mechanism in two
stages. The  first stage is between the three active neutrinos and
their counter parts, the three right handed inert neutrinos $ \nu_R$
$\equiv$ ($\mid \nu_{eR} \rangle$, $\mid \nu_{\mu R} \rangle$, $\mid
\nu_{\tau R} \rangle$). The second stage  is between $N_L$ $\equiv$
$(N_1, N_2, N_3)_L$, $N_R$ $\equiv$ $(N_1, N_2, N_3)_R$. These two
stages are achieved by constraining the elements of
the sub mass matrices in ${\cal M}_{\nu}$ to satisfy the seesaw conditions,
\be
M , M_{RR} \gg \Delta, \Delta^{'}, \Delta^{''}, M_{LL}, \Delta_{LR}, m_{RR}.
\ee
 The  light
neutrino masses for the three active states are given by
\be
M_{\nu_L}\ \simeq\ \,-\Delta M^{-1} \Delta^{\sf T} .
\ee
The active states mix through the matrix elements of $\epsilon =\Delta
M^{-1}$. These mixings are
responsible for the observed solar, atmospheric and reactor neutrino
oscillations. Similarly,
the masses of the light steriles are given by
\be
M_{N_L}\ \simeq\ M_{LL} - \,\Delta_{LR} M^{-1}_{RR} \Delta_{LR}^{\sf T},
\ee
and the states mix via the matrix elements of $\delta =\Delta_{LR}
M^{-1}_{RR}$. Further mixing between the three active states and the
three sterile states are achieved through the off diagonal matrices $
\Delta^{''}$. These mixings are responsible for addressing the reactor
anomaly, the flux
depletion at IceCube and dark matter.
 \\
 %%%%%%%%%%%%%%%%%%%%%%%%%%%%%%%%%%%%%%%%%%%%
 %%                   MODEL B              %%
 %%%%%%%%%%%%%%%%%%%%%%%%%%%%%%%%%%%%%%%%%%%%
\noindent \textbf{Model B:} In this model the light mass eigenstates
are the active states $ \nu_L $ $\equiv$ ($\mid \nu_{eL} \rangle$,
$\mid \nu_{\mu L} \rangle$, $\mid \nu_{\tau L} \rangle$) and their
counter parts, the three right handed inert neutrinos $ \nu_R$
$\equiv$ ($\mid \nu_{eR} \rangle$, $\mid \nu_{\mu R} \rangle$, $\mid
\nu_{\tau R} \rangle$).  In this model also the lightness of the
states is achieved by invoking the seesaw mechanism in two stages. The
first stage is between the three active neutrinos and $N_R$ $\equiv$
$(N_1, N_2, N_3)_R$. The second stage  is between $ \nu_R$ $\equiv$
($\mid \nu_{eR} \rangle$, $\mid \nu_{\mu R} \rangle$, $\mid \nu_{\tau
  R} \rangle$) and $N_L$ $\equiv$ $(N_1, N_2, N_3)_L$.
These two stages are achieved by constraining the elements of the sub
mass matrices in ${\cal M}_{\nu}$ to satisfy the seesaw conditions,
\be
M_{RR},  M_{LL} \gg  M, \Delta, \Delta^{'}, \Delta^{''}, \Delta_{LR}, m_{RR}.
\ee
 The  light
neutrino masses for the three active states are given by
\be
M_{\nu_L}\ \simeq\ \,-\Delta^{'} M_{RR}^{-1} \Delta^{' \sf T}\; .
\ee
The active states mix through the matrix elements of $\epsilon^{'}
=\Delta^{'} M_{RR}^{-1}$. These mixing
matrix elements are responsible for the observed solar, atmospheric
and reactor neutrino oscillations.
Similarly, the masses of the light $ \nu_R$'s are given by
\be
M_{ \nu_R}\ \simeq\ M - \,\Delta^{''} M^{-1}_{LL} \Delta^{'' \sf T},
\ee
and the states mix via the matrix elements of $\delta =\Delta^{''}
M^{-1}_{LL}$. Further mixing between the three
active states and the three sterile states are achieved through the
off diagonal matrices $ \Delta^{''}$.
These mixings are responsible for addressing the reactor anomaly, the
flux depletion at IceCube and dark matter. 
This model also offers the possibility of constructing a pseudo-Dirac
particle\cite{Wolfenstein:1981kw,Petcov:1982ya,Bilenky:1983wt,Kobayashi:2000md} by combining 
two almost degenerate mass eigenstates, one from the active neutrinos
and another  from their right handed counterparts.

\section{High Energy neutrino oscillation}

In the (3+3) model, the  vacuum oscillation probability for the process
$\nu_{\alpha}\rightarrow \nu_{\beta}$ is given as
\bary
P_{\alpha\beta}(L)&=&\delta_{\alpha\beta} -4\sum_{i>j} {\Re} [U^*_{\alpha
  i}U_{\beta i} U_{\alpha j} U^*_{\beta j}] \sin^2 \left (\frac{\Delta
  m^2_{ij}L}{4 E_{\nu}}\right )\nonumber\\
&+& 2 \sum_{i>j} {\Im} [U^*_{\alpha
  i}U_{\beta i} U_{\alpha j} U^*_{\beta j}] \sin \left (\frac{\Delta
  m^2_{ij}L}{2 E_{\nu}}\right ).
\label{prob6}
\eary
where $(i,j=1\, {\rm to} \,6)$ and  15 different
$\Delta m^2_{ij}=m^2_i-m^2_j$ for non-zero and non-degenerate cases. For given $\Delta m^2$, the oscillation probability depends on
the neutrino energy $E_{\nu}$ and the propagation distance (baseline) $L$.
Because CP violation in the neutrino sector
is not observed yet, we take all the phases to be zero and this makes
the U matrix real and  simplifies the oscillation probability to the
following form
\be
P_{\alpha\beta}(L)=\delta_{\alpha\beta} -4\sum_{i>j}  [U_{\alpha
  i}U_{\beta i} U_{\alpha j} U_{\beta j}] \sin^2 \left (\frac{\pi
    L}{L_{\rm osc}}\right ),
\label{prob}
\ee
where $L_{\rm osc}=4\pi E_{\nu}/\Delta m^2_{ij}$ is the oscillation length.
The maximum flavor conversion in the vacuum can take place when
$L=L_{\rm osc}/2$.
If $L\gg L_{\rm osc}$, oscillation are very rapid
and the oscillating term averages out to $1/2$. In this case the
oscillation probability neither depends on neutrino energy $E_{\nu}$ nor on the
distance $L$ from the source. On the other hand
if $L\ll L_{\rm osc}$, the baseline is too short for neutrinos to
oscillate.

In order to explain the solar and atmospheric neutrino oscillation data we take
$\Delta m^2_{21}$ and $\Delta m^2_{31}$ as given
in Eq.(\ref{dm2}) and their corresponding mixing angles.
To explain SBL anomaly we adopt the (3+1) model\cite{Giunti:2011gz}.
In the (3+1) scenario, neutrino masses consist of three  active neutrinos with masses
$m_1,\,m_2$ and $m_3$ that accommodate the observed solar and
atmospheric oscillations and a
sterile state with mass $m_j, (j=4 \,\, \text{or} \,\,  5)$ separated from the active states by
$\Delta m^2_{j1} \sim 1\,{\rm eV}^2\gg \Delta m^2_{21,31}$.
The small squared-mass differences
$\Delta m^2_{21}$ and $\Delta m^2_{31}$  which are responsible for
solar and atmospheric neutrino oscillations respectively have
negligible effects in SBL oscillations. On the other hand,
due to the large $\Delta m^2_{j1}$
and small active-sterile mixing, the effect of the sterile
neutrino on the solar neutrino oscillation and conventional atmospheric neutrino oscillation
($E_{\nu}\sim$ GeV) are also negligible. However, the new large mass-squared
difference induces an
active-sterile oscillation at short base-lines $\sim10$ m for neutrinos with energy $E_{\nu}\sim 100$ MeV, which
is invoked to interpret the SBL anomaly.

 In order to explain the depletion of the high energy neutrino flux in  IceCube
we assume that the sterile neutrino $\mid \nu_{4} \rangle$ or $\mid \nu_{5} \rangle$  with mass  $m_4$ or $m_5$ and  which does not participate in the SBL
oscillation will be almost degenerate in mass with $\mid \nu_{1} \rangle$ or $\mid \nu_{2} \rangle$ and we can
estimate its value $\Delta m^2\simeq 4.0 \times 10^{-16}\, {\rm eV}^2$
in the proceeding section for maximum flavor
conversion on Earth.
%Also if we neglect the reactor anomaly then in
%this scenario both $m_4$ and $m_5$ will be degenerate with $m_1$ and/or $m_2$.
This gives many possibilities for $\Delta m^2$ to be considered and we
take into account many of them in our analysis as shown in Table-I. 
A sterile neutrino with a mass of
several (1-10) keV is a viable candidate for dark matter\cite{Dodelson:1993je},  can 
explain the pulsar kicks\cite{Kusenko:1997sp} and can also play a role in other astrophysical phenomena.
Finally to be able to explain the DM
problem, we assume that the sixth neutrino mass eigenstate has mass
$m_6\simeq 1$ keV and is almost decoupled from the rest of the
neutrinos, both active and sterile. 
%%%%%%%%%%%%%%%%%%%%%%%%%%
%%%%%%%%%%%%%%%%%%%%%%%%%%
\section{The mixing matrix}
%%%%%%%%%%%%%%%%%%%%%%%%%%
%%%%%%%%%%%%%%%%%%%%%%%%%%

The matrix  U in Eq.(\ref{nustate}) is a unitary $6\times
6$ matrix and in general can be parameterized
by 15 real angles and 10 Dirac phases entering directly in the
mixing matrix. The remaining 5 phases enter as a diagonal matrix and
sits outside the matrix. The only mixing angles which are measured
experimentally are  $\theta_{12}$,
$\theta_{23}$ and $\theta_{13}$. In discussing physics beyond the Standard
Model scenario one has to incorporate these measured
parameters in the analysis pertaining to oscillations  involving sterile neutrinos. Models involving one (3+1), two (3+2) and three
(3+3)\cite{Giunti:2011gz,Archidiacono:2012ri,Donini:2012tt,Keranen:2003xd,Biermann:2006bu,
Conrad:2012qt,Dev:2012bd,Merle:2013gea,Pakvasa:2012db}, sterile neutrinos have
been proposed to explain the discussed discrepancies where many simple
parametrization of the matrix $U$ have been used \cite{Keranen:2003xd,Gupta:2013vva}.

%In order to be able to explain the neutrino oscillation data between the active
%neutrinos and the above three problems,

 In order to address the aforementioned problems and at the same time accommodate existing data on the observed oscillations between the active neutrinos we propose the following $6\times 6$ form for extended PMNS matrix $U$,
\bwt
\renewcommand{\arraystretch}{2}
\newcommand*{\temp}{\multicolumn{1}{r|}{}}
\be
%$$
U\simeq \left(\begin{array}{ccccccc}
0.824 &0.515 & 0.136 &\temp & 0.138 & 0.139 & 1.0\times 10^{-3}\\
-0.501 & 0.527 & 0.583 &\temp & 0.243 & 0.203 & 0.174 \\
0.244 & -0.670 & 0.629 &\temp & 0.223 & 0.197 & 0.086 \\ \cline{1-7}
-0.052 & 0.070 & -0.409 &\temp & 0.901 & 0.078 & 0.086\\
-0.050 & -0.047 & -0.261 &\temp & -0.214 & 0.935 & 0.085\\
0.076 & -0.025 & -0.101 &\temp & -0.124 & -0.142 & 0.974\\
\end{array}\right).
%$$
\label{sixmat}
\ee
\ewt
Notice that the first $3\times 3$ block
diagonal entries are responsible for explaining the solar, atmospheric
and the reactor neutrino data and all the mixing matrix elements in this
block diagonal are compatible with the constraints given by
experiments. The active ($\nu_e$, $\nu_{\mu}$ and $\nu_{\tau}$ )
content of the three additional mass eigenstates has to be small,
which is shown in the first $3\times 3$ off diagonal block in
Eq.(\ref{sixmat}). Unitarity of $U$ implies the following constraints on the remaining matris elements,
\be
X_i\equiv\sum_{\alpha=e,\mu,\tau} |U_{\alpha i}|^2 \le 0.3,
\ee
for each $i=4-6$.  Similarly for each $\alpha=e,\mu,\tau$
\be
X_{\alpha}\equiv\sum_{i=4-6} |U_{\alpha i}|^2 \le 0.3.
\ee
Our extended matrix gives $0.04 \le X_{i} \le 0.13$ for $i=4-6$ and
$ 0.04\le X_{\alpha} \le 0.13$ for  $\alpha=e,\mu,\tau$.

To further tighten the constraint on the active sterile mixing
we can use the effective neutrino mass in $\beta$ decay experiments,
which is given by
\be
m_{e}=\sqrt { \left ( \sum_i | U_{ei}|^2 m^2_i  \right)}.
\ee
This contribution gives the distortion of the electron energy spectrum due to
nonzero neutrino mass and mixing and the current bound on this parameter
is $m_{e}\le 2.2$ eV\cite{Kraus:2004zw}. Similarly  in the neutrinoless double beta
decay experiments the effective neutrino mass parameter is given by
\be
\langle m \rangle_{ee} = \mid \sum_i U^2_{ei} m_i \mid.
\ee
The current bound on this parameter is $\langle m \rangle_{ee}
< 0.26$ eV\cite{Arnold:2005rz,Bongrand:2011ei}. In our analysis, we have three different mass scales, one scale is
in the  sub-eV range and can even be smaller,  making  $m_1$ degenerate with
$m_2$, another scale is of order eV, corresponding to either $m_4$ or $m_5$ and  the third one
is the keV scale corresponding to $m_6$. The effective neutrino
mass parameter in both the experiments has to get contribution mainly from keV and eV mass
eigenstates and to obtain these constraint, we must have
$|U_{e6}|\lesssim 10^{-3}$.
Our extended U matrix satisfies this characteristic i.e. $m_e \le
1.05$ eV and $\langle m \rangle_{ee} \le 10^{-3}$ eV.
Also, to preserve the well known mixing between the active
neutrinos, the mixing between the active and the sterile neutrinos are
required to be small.
%Typically their values ranges from $\sim
%10^{-3}$ to $\sim 10^{-1}$, as can be inferred from Eq.(\ref{sixmat}).
Thus the flavor mixing  matrix elements of the active neutrinos in $U$ (first diagonal block of
Eq.(\ref{sixmat})) constraint the remaining mixing elements between the active
and the sterile neutrinos to be small but sizable,  translating into
the corresponding  mixing angles to be a  few degrees at most
($\theta_{ij}\le 15^{\rm o}$ for $i,j$ from 4 to 6).

%%%%%%%%%%%%%%%%%%%%%%%%%%
%%%%%%%%%%%%%%%%%%%%%%%%%%
\section{high energy astrophysical neutrinos}
%%%%%%%%%%%%%%%%%%%%%%%%%%
%%%%%%%%%%%%%%%%%%%%%%%%%%

It is believed that, GRBs which are about 100 Mpc away from us
are the sources of UHECRs with energies above $10^{18}$ eV\cite{Halzen:2002pg,Becker:2007sv,Murase:2008mr}.
In the fireball scenario of the GRB emission\cite{Murase:2005hy,Fox:2006iu},
protons are Fermi accelerated to ultra-high energy and constitute
probably part of the UHECRs that we observe on Earth.  The deep
inelastic collision of
these high energy protons with the expanding shock wave as well as
with the surrounding background can produce charged and neutral
pions. While the decay of neutral pion can give high energy gamma-rays,
decay of charged pions will produce high energy neutrinos. So there is
some correlation among the UHECRs, high energy gamma-rays and the high
energy neutrinos.

The conventional wisdom is that at the source the flux
ratio is $\Phi^0_{\nu_e}:\Phi^0_{\nu_{\mu}}:\Phi^0_{\nu_{\tau}}=1:2:0$
( $\Phi^0_{\nu_{\alpha}}$ is the sum of neutrino and anti-neutrino
fluxes for the flavor $\alpha$ at the source) due to the decay of charged
pions. The vacuum oscillation
of these neutrinos on their way to Earth would average out to  the observed ratio
($1:1:1$)\cite{Learned:1994wg}. For high energy neutrinos above $\sim 1$ PeV, muon energy is
degraded in strong magnetic field or get absorbed in the stellar
medium. So high energy muon neutrinos will be absent and the flux
ratio at the source is
modified to ($0:1:0$) \cite{Rachen:1998fd,Lipari:2007su,Pakvasa:2007dc}. This will be further modified to ($1:1.8:1.8$) at
Earth after vacuum oscillation\cite{Kashti:2005qa}. 

Neutron beta-decay will also contribute to the neutrino flux ratio.
Being neutral, neutrons can not be accelerated directly by
the GRB jet. So these neutrons have to be produced as
secondaries. Around the  GRB environment, high energy neutrons can be
produced through the following channels:
interaction of Fermi accelerated high energy protons in the GRB jet
with the ambient hydrogen ($pp$), dissociation of accelerated ions
($A$) by colliding with the ambient hydrogen ($Ap$), interaction of high
energy protons with the ambient photons ($p\gamma$) and photodissociation of
accelerated ions ($A\gamma$)\cite{Crocker:2004bb}. These high energy secondary neutrons will decay in flight and
produce $\bar{\nu}_e$ which will give a flux ratio ($1:0:0 $)\cite{Rachen:1998fd,Lipari:2007su,Pakvasa:2007dc}.
However, this scenario has at least one
shortcoming. 
%The first one is that the secondary neutron has to be
%extremely energetic $E_n\sim 10^3 E_{\nu}$, which is probably, not so
%easy to achieve and the second problem is that, 
In the GRB  environment along with these neutrons, high
energy pions are also produced. The high energy charged pions will
decay to high energy neutrinos and their energy will be order of
magnitude higher than the $\bar{\nu_e}$ energy produced in neutron
beta decay. Also the neutrino flux from pion decay will be higher than
the one from the neutron decay. 
So in a astrophysical environment, a pure neutron source having
the flux ratio ($1:0:0 $) is highly unrealistic. 

 Based on the observed flux of
UHECRs, Waxman and Bahcall estimated the neutrino flux, which is
$
E^2_{\nu} dN_{\nu}/dE_{\nu} \sim 5\times 10^{-9}\, {\rm GeV}{\rm
  cm}^{-2}{\rm s^{-1}}{\rm sr^{-1}}
$
in the energy range $\sim$ 100 TeV - 10 PeV\cite{Waxman:1997ti}.
For GRBs at a redshift of $z\sim 1$ and $L\sim 100$ Mpc with
neutrinos energy $100\,{\rm TeV} \lesssim E_{\nu}\lesssim 10$ PeV, the maximum flavor conversion will
take place for
\be
4.0\times 10^{-17}\, {\rm eV}^2 \lesssim \Delta m^2 \lesssim 4.0\times 10^{-15}\,{\rm eV}^2.
\ee
In other words, the high energy GRB neutrinos can not probe mass
squared difference smaller than $\Delta
m^2\simeq 4.0\times 10^{-17}\, {\rm eV}^2$.
For our estimate of the neutrino flux we will use this result for
the maximum conversion of neutrinos in the IceCube detector.
The oscillation length for standard neutrinos  as well as for
neutrinos satisfying $\Delta m^2 \sim 1\, {\rm eV}^2$
and $\Delta m^2 \sim 1\, {\rm keV}^2$ are very short compared
to the astrophysical distances which corresponds to the  condition
$L\gg L_{\rm osc}$ and the oscillation probability will
be averaged out for these cases which will be independent of the
neutrino energy and the distance from the source. For our analysis,
here we consider the neutrino energy $E_{\nu}=1$ PeV which
corresponds to $\Delta m^2\simeq 4.0\times 10^{-16}\, {\rm eV}^2$ for
maximum flavor conversion on Earth and for this case we replace the
oscillatory factor in Eq.(\ref{prob}) by unity.

For the treatment of the neutrino oscillation in Eq.(\ref{prob}), we
have neglected  the matter effects for  GRBs as well as the Earth. The
reasons are twofold. (1) The region of the GRB fireball
where these high energy neutrinos are produced,  has very low matter
density which makes the matter potential contribution negligible. (2) For
$\Delta m^2\simeq 4.0 \times 10^{-16}\, {\rm eV}^2$ the
average  potential experienced by a PeV neutrino inside the Earth is
$\sqrt{2} G_F n_e\gg \Delta m^2/2 E_{\nu}$. Thus also, the Earth's matter has negligible effect on
neutrino oscillation. 

%%%%%%%%%%%%%%%%%%%%%%%%%%
%%%%%%%%%%%%%%%%%%%%%%%%%%
\section{Results and discussion}
%%%%%%%%%%%%%%%%%%%%%%%%%%
%%%%%%%%%%%%%%%%%%%%%%%%%%

In the light of insufficient
detailed  knowledge on the region of the GRB fireball and the region
surrounding it where the
high energy neutrinos are produced as discussed in the previous section, we consider three different flux
ratios at the source: the conventional one, ($1:2:0$),  the muon-damped
source, ($0:1:0$) and the beta beam ($1:0:0$) . Of
course this is not the only uncertainty that affects  the flux
determination on Earth, there is also uncertainty in the elements of
the U matrix and a number of other astrophysical factors:
the shape of the neutrino spectra depends on the primary cosmic rays
energy spectra and of the target material and at very high energy,
semileptonic decay of the charm quarks will give rise to extra neutrinos.
In our analysis,  we neglect the last two uncertainties in
calculating  the flux ration on Earth.

After travelling a distance $L$, the neutrino flux of a given
flavor on Earth is given by
\be
\Phi_{\nu_{\alpha}}=\sum_{\beta} P_{\alpha\beta} \Phi^0_{\nu_{\beta}}.
\ee
%In principle we should evaluate the factor $\sin^2(\pi L/L_{\rm osc})$
%instead of replacing by unity, because this depends on $L$ and $E_{\nu}$.
The condition
$L\gg L_{\rm osc}$ is satisfied for the standard neutrinos as well as
for neutrinos satisfying $\Delta m^2 \sim 1\, {\rm eV}^2$ and $\Delta m^2 \sim 1\, {\rm keV}^2$. For all
these cases the oscillatory term in Eq.(\ref{prob}) will be replaced
by a factor 1/2. We keep the sixth neutrino mass $m_6=1$ keV fixed
throughout the calculation.
Table-I summarizes our findings. We have considered six different
possibilities that give sizable very high energy neutrino flux
depletion in IceCube. In first four cases we have taken either $m_4$
or $m_5$ $\sim 1$ eV. For the remaining cases both $m_4$ and $m_5$ are
almost degenerate with either $m_1$ or $m_2$ but not both. In these
last two cases considered, we do not have $\sim 1$ eV neutrinos mass
and will be unable to explain the SBL anomaly.

%\bwt
\begin{table}[ht]
\centering\begin{tabular}{|c|c|c|c|c|}
\hline
Possibility&$\Delta m^2_{ij}$ &
$1:2:0$ & $0:1:0$ & $1:0:0$\\[1ex]
\hline
\hline
I&$\Delta m^2_{41}$ & 1.07:0.75:0.74\,\,\, & 0.28:0.23:0.29\,\,\, & 0.51:0.28:0.16\,\,\,\\
II&$\Delta m^2_{51}$ & 1.06:0.76:0.74&  0.28:0.24:0.29 & 0.51:0.28:0.16 \\
\hline
III&$\Delta m^2_{42}$ &0.99:0.69:0.82 &  0.23:0.23:0.32 & 0.52:0.23:0.19\\
IV&$\Delta m^2_{52}$ & 0.99:0.72:0.80 &  0.24:0.24:0.31 & 0.52:0.24:0.19\\
\hline
V&$\Delta m^2_{41,51}$ & 1.08:0.72:0.74& 0.30:0.21:0.30 & 0.48:0.30:0.14\\
VI&$\Delta m^2_{42,52}$ & 0.94:0.62:0.89 & 0.22:0.20:0.34 & 0.51:0.22:0.21\\
%\hline
%V&$\Delta m^2_{41},\, \Delta m^2_{51}$ & - & 1.06:0.90:0.86
%& 0.29:0.30:0.34\\
%VI&$\Delta m^2_{42},\, \Delta m^2_{52}$ &- & 1.0:0.85:0.92
%& 0.25:0.30:0.36\\
%VII&$\Delta m^2_{41},\, \Delta m^2_{52}$ &- & 1.04:0.88:0.89
%& 0.27:0.30:0.35\\[1ex]
\hline
\end{tabular}
\caption{The first column gives different possibilities that we have
  considered. In the second column the mass splittings of the sterile
  neutrino with the active one are given where we fix 
$\Delta m^2_{ij}\simeq 4.0\times 10^{-16}\,{\rm eV}^2$ for maximum
flavor conversion on Earth.
The  third, the fourth and the fifth columns give the flux ratio on Earth by considering the flux ratio
  at the source to be  $(1:2:0)$, ($0:1:0$) and ($1:0:0 $) respectively.
}
\end{table}
%\ewt

As shown in Table-I, for the initial neutrino flux ratio  ($1:2:0$), we observe that $\Phi_{\nu_e} > \Phi_{\nu_{\mu}}$,
$\Phi_{\nu_{\tau}}$ is satisfied always. For the mass degeneracy involving $m_1$,
the electron neutrino flux $\Phi_{\nu_e}$ , is always enhanced from
its vacuum value by 6\% to 8\%. At the same time 
$\Phi_{\nu_{\mu}}$decreases by 24\% to
28\% and $\Phi_{\nu_{\tau}}$  by 26\%. 
On the other hand,  for the mass degeneracy
involving $m_2$,  while the $\Phi_{\nu_e}$ is decreased by 1\% to 6\%,
$\Phi_{\nu_{\mu}}$ is decreased substantially by 28\% to 38\% and
$\Phi_{\nu_{\tau}}$  is decreased by 11\% to 20\%. 
The substantial
depletion in  $\Phi_{\nu_{\mu}}$ and 
$\Phi_{\nu_{\tau}}$  in  ($1:2:0$) scenario is due to the sizable
mixing of the $\nu_{\mu}$ and $\nu_{\tau}$ with the sterile neutrinos,
which can be seen from the mixing matrix given in Eq.(\ref{sixmat}).

For the initial
flux ratio ($0:1:0$) as shown in Table-I,  the observed flux on Earth is almost identical
for mass degeneracies involving $m_1$ (possibilities I and II). Similarly
it is also almost identical for degeneracies involving $m_2$ (possibilities III and IV). In
all these cases we find
$\Phi_{\nu_{\tau}}>\Phi_{\nu_{\mu}}>\Phi_{\nu_e}$ but still lower than
the vacuum oscillation value of 1/3. In the last two cases
(possibilities V and VII), the steriles neutrinos with masses $m_4$ and $m_5$ are
taken to be degenerate with either $m_1$ or $m_2$. We see that the
muon neutrino flux is degraded substantially by as much as 38\%.
In general our results show that there is
marked decrease in the muon neutrino flux.

For the beta beam flux ratio ($1:0:0$), as shown in last column of
Table-I, we observe that $\Phi_{\nu_e}
> \Phi_{\nu_{\mu}} > \Phi_{\nu_{\tau}}$ is always satisfied and in all
these cases while the tau neutrino flux is heavily suppressed (36\% to
43\%), the  $\Phi_{\nu_e}$ is dramatically increased by as much as
45\% to 58\%. The $\Phi_{\nu_{\mu}}$ is depleted by 9\% to 34\%.
We observe a substantial depletion in muon neutrino flux in the above three
scenarios which can clearly be  measurable by IceCube.

IceCube can isolate the   muon neutrino events from the rest
through the track and shower events.  
The most probable signature of the sterile neutrinos is the depletion of
muon neutrino flux due to its mixing with the former. In all the three 
cases, ($1:2:0$), ($1:0:0$) and ($0:1:0$), we obtain
depletion in the muon neutrino flux as well as in tau neutrino flux.
It has also been
argued that due to active-sterile flavor  mixing, there will be an excess of electron
neutrinos with a particular energy and zenith angle
dependence\cite{Halzen:2011yq}. For conventional flux ratio ($1:2:0$)
and beta beam flux ratio
($1:0:0$)  we get an
excess of electron neutrino flux on Earth.
In the conventional scenario this excess in $\Phi_{\nu_e}$ is due to
the mixing of
neutrino of mass $m_1$ with the steriles of mass $m_4$ and $m_5$, which
is absent in the muon-damped scenario ($0:1:0$). 
But in the beta beam
scenario, the enhancement in $\Phi_{\nu_e}$ is very high (between 45\% to 58\%) for
all active-sterile mixing.

For SBL neutrino oscillation, except for the  $\Delta m^2\sim 1\, {\rm eV}^2$
term,  no other mass square difference term will contribute. The
oscillatory term with
$\Delta m^2_{6i} L/4 E_{\nu} \gg 1$ for $i=1-5$ ($L\sim 10$
m and $E_{\nu}\sim 100$ MeV) will, in principle, contribute to the SBL anomaly. But
the mixing of the sixth massive neutrino is vanishingly small resulting in
a contribution that is  negligible. Thus the  sixth neutrino
decouples from the rest and serves as the nonbaryonic dark matter of the Universe\cite{Kusenko:2009up,Boyarsky:2009ix}.
There are also other explanations for the non-observation
of these high energy neutrinos in IceCube, where it is argued that, GRB may not be the source
of  high energy cosmic rays, in that case there will be no neutrinos.
 Another explanation is that the GRB fireball calculation
of the neutrino flux is subject to sufficiently large astrophysical
ambiguities to escape the IceCube limit\cite{Hummer:2011ms}. 

%%%%%%%%%%%%%%%%%%%%%%%%%%%%%%%%%%%%%%%%%%%%%%%%%%%%%%%%%%%%%%%%%%%%%%%%%%%%%%%%%%%%%%%%%%%%

%%%%%%%%%%%%%%%%%%%%%%%%%%%%%%%%%%%%%%%%%%%%%%%%%%%%%%%%%%%%%%%%%%%%%%%%%%%%%%%%%%%%%%%%%%%
We thank S. Mohanty and S. Pakvasa for valuable
comments and discussions. The  work of S.S. is partially
supported by DGAPA-UNAM (Mexico) Project
No. IN103812 and the work of S.R. is partially supported by the
Department of Energy (USA) grant No. DE-SC0005366/001.


\begin{thebibliography}{55}
\expandafter\ifx\csname natexlab\endcsname\relax\def\natexlab#1{#1}\fi
\expandafter\ifx\csname bibnamefont\endcsname\relax
  \def\bibnamefont#1{#1}\fi
\expandafter\ifx\csname bibfnamefont\endcsname\relax
  \def\bibfnamefont#1{#1}\fi
\expandafter\ifx\csname citenamefont\endcsname\relax
  \def\citenamefont#1{#1}\fi
\expandafter\ifx\csname url\endcsname\relax
  \def\url#1{\texttt{#1}}\fi
\expandafter\ifx\csname urlprefix\endcsname\relax\def\urlprefix{URL }\fi
\providecommand{\bibinfo}[2]{#2}
\providecommand{\eprint}[2][]{\url{#2}}


%\cite{Abe:2010hy}
\bibitem{Abe:2010hy}
  K.~Abe {\it et al.}  [Super-Kamiokande Collaboration],
  %``Solar neutrino results in Super-Kamiokande-III,''
  Phys.\ Rev.\ D {\bf 83}, 052010 (2011)
  [arXiv:1010.0118 [hep-ex]].
  %%CITATION = ARXIV:1010.0118;%%
  %78 citations counted in INSPIRE as of 26 Feb 2013

%\cite{Adamson:2011ig}
\bibitem{Adamson:2011ig}
  P.~Adamson {\it et al.}  [MINOS Collaboration],
  %``Measurement of the neutrino mass splitting and flavor mixing by MINOS,''
  Phys.\ Rev.\ Lett.\  {\bf 106}, 181801 (2011)
  [arXiv:1103.0340 [hep-ex]].
  %%CITATION = ARXIV:1103.0340;%%
  %127 citations counted in INSPIRE as of 26 Feb 2013

%\cite{Ahmad:2002jz}
\bibitem{Ahmad:2002jz}
  Q.~R.~Ahmad {\it et al.}  [SNO Collaboration],
  %``Direct evidence for neutrino flavor transformation from neutral current interactions in the Sudbury Neutrino Observatory,''
  Phys.\ Rev.\ Lett.\  {\bf 89}, 011301 (2002)
  [nucl-ex/0204008].
  %%CITATION = NUCL-EX/0204008;%%

%\cite{Abe:2008aa}
\bibitem{Abe:2008aa}
  S.~Abe {\it et al.}  [KamLAND Collaboration],
  %``Precision Measurement of Neutrino Oscillation Parameters with KamLAND,''
  Phys.\ Rev.\ Lett.\  {\bf 100}, 221803 (2008)
  [arXiv:0801.4589 [hep-ex]].
  %%CITATION = ARXIV:0801.4589;%%

%\cite{Fukuda:1998mi}
\bibitem{Fukuda:1998mi}
  Y.~Fukuda {\it et al.}  [Super-Kamiokande Collaboration],
  %``Evidence for oscillation of atmospheric neutrinos,''
  Phys.\ Rev.\ Lett.\  {\bf 81}, 1562 (1998)
  [hep-ex/9807003].
  %%CITATION = HEP-EX/9807003;%%

%\cite{Michael:2006rx}
\bibitem{Michael:2006rx}
  D.~G.~Michael {\it et al.}  [MINOS Collaboration],
  %``Observation of muon neutrino disappearance with the MINOS detectors and the NuMI neutrino beam,''
  Phys.\ Rev.\ Lett.\  {\bf 97}, 191801 (2006)
  [hep-ex/0607088].
  %%CITATION = HEP-EX/0607088;%%

%\cite{Abe:2011fz}
\bibitem{Abe:2011fz}
  Y.~Abe {\it et al.}  [DOUBLE-CHOOZ Collaboration],
  %``Indication for the disappearance of reactor electron antineutrinos in the Double Chooz experiment,''
  Phys.\ Rev.\ Lett.\  {\bf 108}, 131801 (2012)
  [arXiv:1112.6353 [hep-ex]].
  %%CITATION = ARXIV:1112.6353;%%

%\cite{Ahn:2012nd}
\bibitem{Ahn:2012nd}
  J.~K.~Ahn {\it et al.}  [RENO Collaboration],
  %``Observation of Reactor Electron Antineutrino Disappearance in the RENO Experiment,''
  Phys.\ Rev.\ Lett.\  {\bf 108}, 191802 (2012)
  [arXiv:1204.0626 [hep-ex]].
  %%CITATION = ARXIV:1204.0626;%%

%\cite{An:2012eh}
\bibitem{An:2012eh}
  F.~P.~An {\it et al.}  [DAYA-BAY Collaboration],
  %``Observation of electron-antineutrino disappearance at Daya Bay,''
  Phys.\ Rev.\ Lett.\  {\bf 108}, 171803 (2012)
  [arXiv:1203.1669 [hep-ex]].
  %%CITATION = ARXIV:1203.1669;%%

%\cite{Aguilar:2001ty}
\bibitem{Aguilar:2001ty}
  A.~Aguilar-Arevalo {\it et al.}  [LSND Collaboration],
  %``Evidence for neutrino oscillations from the observation of anti-neutrino(electron) appearance in a anti-neutrino(muon) beam,''
  Phys.\ Rev.\ D {\bf 64}, 112007 (2001)
  [hep-ex/0104049].
  %%CITATION = HEP-EX/0104049;%%
  %943 citations counted in INSPIRE as of 26 Feb 2013

%\cite{AguilarArevalo:2010wv}
\bibitem{AguilarArevalo:2010wv}
  A.~A.~Aguilar-Arevalo {\it et al.}  [MiniBooNE Collaboration],
  %``Event Excess in the MiniBooNE Search for $\bar \nu_\mu \rightarrow \bar \nu_e$ Oscillations,''
  Phys.\ Rev.\ Lett.\  {\bf 105}, 181801 (2010)
  [arXiv:1007.1150 [hep-ex]].
  %%CITATION = ARXIV:1007.1150;%%
  %230 citations counted in INSPIRE as of 26 Feb 2013

%\cite{Mention:2011rk}
\bibitem{Mention:2011rk}
  G.~Mention, M.~Fechner, T.~.Lasserre, T.~.A.~Mueller, D.~Lhuillier, M.~Cribier and A.~Letourneau,
  %``The Reactor Antineutrino Anomaly,''
  Phys.\ Rev.\ D {\bf 83}, 073006 (2011)
  [arXiv:1101.2755 [hep-ex]].
  %%CITATION = ARXIV:1101.2755;%%
  %167 citations counted in INSPIRE as of 26 Feb 2013

%\cite{ALEPH:2005ab}
\bibitem{ALEPH:2005ab}
  S.~Schael {\it et al.}  [ALEPH and DELPHI and L3 and OPAL and SLD
  and LEP Electroweak Working Group and SLD Electroweak Group and SLD
  Heavy Flavour Group Collaborations],
  %``Precision electroweak measurements on the $Z$ resonance,''
  Phys.\ Rept.\  {\bf 427}, 257 (2006)
  [hep-ex/0509008].
  %%CITATION = HEP-EX/0509008;%%
  %758 citations counted in INSPIRE as of 26 Feb 2013



%\cite{Kusenko:2009up}
\bibitem{Kusenko:2009up}
  A.~Kusenko,
  %``Sterile neutrinos: The Dark side of the light fermions,''
  Phys.\ Rept.\  {\bf 481}, 1 (2009)
  [arXiv:0906.2968 [hep-ph]] and references therein.
  %%CITATION = ARXIV:0906.2968;%%
  %98 citations counted in INSPIRE as of 25 Feb 2013

%\cite{Boyarsky:2009ix}
\bibitem{Boyarsky:2009ix}
  A.~Boyarsky, O.~Ruchayskiy and M.~Shaposhnikov,
  %``The Role of sterile neutrinos in cosmology and astrophysics,''
  Ann.\ Rev.\ Nucl.\ Part.\ Sci.\  {\bf 59}, 191 (2009)
  [arXiv:0901.0011 [hep-ph]].
  %%CITATION = ARXIV:0901.0011;%%
  %123 citations counted in INSPIRE as of 25 Feb 2013


%\cite{Fuller:2003gy}
\bibitem{Fuller:2003gy}
  G.~M.~Fuller, A.~Kusenko, I.~Mocioiu and S.~Pascoli,
  %``Pulsar kicks from a dark-matter sterile neutrino,''
  Phys.\ Rev.\ D {\bf 68}, 103002 (2003)
  [astro-ph/0307267].
  %%CITATION = ASTRO-PH/0307267;%%
  %135 citations counted in INSPIRE as of 26 Feb 2013


%\cite{Halzen:2002pg}
\bibitem{Halzen:2002pg}
  F.~Halzen and D.~Hooper,
  %``High-energy neutrino astronomy: The Cosmic ray connection,''
  Rept.\ Prog.\ Phys.\  {\bf 65}, 1025 (2002)
  [astro-ph/0204527].
  %%CITATION = ASTRO-PH/0204527;%%
  %246 citations counted in INSPIRE as of 26 Feb 2013

%\cite{Murase:2008mr}
\bibitem{Murase:2008mr} 
  K.~Murase, K.~Ioka, S.~Nagataki and T.~Nakamura,
  %``High-energy cosmic-ray nuclei from high- and low-luminosity gamma-ray bursts and implications for multi-messenger astronomy,''
  Phys.\ Rev.\ D {\bf 78}, 023005 (2008)
  [arXiv:0801.2861 [astro-ph]].
  %%CITATION = ARXIV:0801.2861;%%
  %65 citations counted in INSPIRE as of 31 Oct 2013


%\cite{Becker:2007sv}
\bibitem{Becker:2007sv}
  J.~K.~Becker,
  %``High-energy neutrinos in the context of multimessenger physics,''
  Phys.\ Rept.\  {\bf 458}, 173 (2008)
  [arXiv:0710.1557 [astro-ph]].
  %%CITATION = ARXIV:0710.1557;%%
  %86 citations counted in INSPIRE as of 26 Feb 2013

%\cite{Murase:2005hy}
\bibitem{Murase:2005hy} 
  K.~Murase and S.~Nagataki,
  %``High energy neutrino emission and neutrino background from gamma-ray bursts in the internal shock model,''
  Phys.\ Rev.\ D {\bf 73}, 063002 (2006)
  [astro-ph/0512275].
  %%CITATION = ASTRO-PH/0512275;%%
  %70 citations counted in INSPIRE as of 31 Oct 2013

%\cite{Halzen:2010yj}
\bibitem{Halzen:2010yj}
  F.~Halzen and S.~R.~Klein,
  %``IceCube: An Instrument for Neutrino Astronomy,''
  Rev.\ Sci.\ Instrum.\  {\bf 81}, 081101 (2010)
  [arXiv:1007.1247 [astro-ph.HE]].
  %%CITATION = ARXIV:1007.1247;%%
  %41 citations counted in INSPIRE as of 26 Feb 2013

%\cite{Abbasi:2012zw}
\bibitem{Abbasi:2012zw}
  R.~Abbasi {\it et al.}  [IceCube Collaboration],
  %``An absence of neutrinos associated with cosmic-ray acceleration in $\gamma$-ray bursts,''
  Nature {\bf 484}, 351 (2012)
  [arXiv:1204.4219 [astro-ph.HE]].
  %%CITATION = ARXIV:1204.4219;%%
  %44 citations counted in INSPIRE as of 26 Feb 2013

%\cite{Abbasi:2011qc}
\bibitem{Abbasi:2011qc}
  R.~Abbasi {\it et al.}  [IceCube Collaboration],
  %``Limits on Neutrino Emission from Gamma-Ray Bursts with the 40 String IceCube Detector,''
  Phys.\ Rev.\ Lett.\  {\bf 106}, 141101 (2011)
  [arXiv:1101.1448 [astro-ph.HE]].
  %%CITATION = ARXIV:1101.1448;%%
  %47 citations counted in INSPIRE as of 26 Feb 2013


%\cite{Waxman:1997ti}
\bibitem{Waxman:1997ti}
  E.~Waxman and J.~N.~Bahcall,
  %``High-energy neutrinos from cosmological gamma-ray burst fireballs,''
  Phys.\ Rev.\ Lett.\  {\bf 78}, 2292 (1997)
  [astro-ph/9701231].
  %%CITATION = ASTRO-PH/9701231;%%
  %551 citations counted in INSPIRE as of 26 Feb 2013

%\cite{Chikashige:1980qk}
\bibitem{Chikashige:1980qk}
  Y.~Chikashige, R.~N.~Mohapatra and R.~D.~Peccei,
  %``Spontaneously Broken Lepton Number and Cosmological Constraints on the Neutrino Mass Spectrum,''
  Phys.\ Rev.\ Lett.\  {\bf 45}, 1926 (1980).
  %%CITATION = PRLTA,45,1926;%%
  %232 citations counted in INSPIRE as of 26 Feb 2013

%\cite{Gelmini:1980re}
\bibitem{Gelmini:1980re}
  G.~B.~Gelmini and M.~Roncadelli,
  %``Left-Handed Neutrino Mass Scale and Spontaneously Broken Lepton Number,''
  Phys.\ Lett.\ B {\bf 99}, 411 (1981).
  %%CITATION = PHLTA,B99,411;%%
  %757 citations counted in INSPIRE as of 26 Feb 2013

%\cite{Beacom:2002vi}
\bibitem{Beacom:2002vi}
  J.~F.~Beacom, N.~F.~Bell, D.~Hooper, S.~Pakvasa and T.~J.~Weiler,
  %``Decay of high-energy astrophysical neutrinos,''
  Phys.\ Rev.\ Lett.\  {\bf 90}, 181301 (2003)
  [hep-ph/0211305].
  %%CITATION = HEP-PH/0211305;%%
  %120 citations counted in INSPIRE as of 26 Feb 2013

%\cite{Li:2011ah}
\bibitem{Li:2011ah} 
  Z.~Li,
  %``Note on the Normalization of Predicted GRB Neutrino Flux,''
  Phys.\ Rev.\ D {\bf 85}, 027301 (2012)
  [arXiv:1112.2240 [astro-ph.HE]].
  %%CITATION = ARXIV:1112.2240;%%
  %19 citations counted in INSPIRE as of 22 Aug 2013

%\cite{Hummer:2011ms}
\bibitem{Hummer:2011ms} 
 S.~Hummer, P.~Baerwald and W.~Winter,
 %``Neutrino Emission from Gamma-Ray Burst Fireballs, Revised,''
 Phys.\ Rev.\ Lett.\  {\bf 108}, 231101 (2012)
 [arXiv:1112.1076 [astro-ph.HE]].
 %%CITATION = ARXIV:1112.1076;%%
 %25 citations counted in INSPIRE as of 10 Jun 2013

%\cite{He:2012tq}
\bibitem{He:2012tq} 
  H.~-N.~He, R.~-Y.~Liu, X.~-Y.~Wang, S.~Nagataki, K.~Murase and Z.~-G.~Dai,
  %``Icecube non-detection of GRBs: Constraints on the fireball properties,''
  Astrophys.\ J.\  {\bf 752}, 29 (2012)
  [arXiv:1204.0857 [astro-ph.HE]].
  %%CITATION = ARXIV:1204.0857;%%
  %27 citations counted in INSPIRE as of 22 Aug 2013


%\cite{Gao:2012ay}
\bibitem{Gao:2012ay} 
  S.~Gao, K.~Asano and P.~Meszaros,
  %``High Energy Neutrinos from Dissipative Photospheric Models of Gamma Ray Bursts,''
  JCAP {\bf 1211}, 058 (2012)
  [arXiv:1210.1186 [astro-ph.HE]].
  %%CITATION = ARXIV:1210.1186;%%
  %4 citations counted in INSPIRE as of 23 Aug 2013

%\cite{Baerwald:2013pu}
\bibitem{Baerwald:2013pu} 
  P.~Baerwald, M.~Bustamante and W.~Winter,
  %``UHECR escape mechanisms for protons and neutrons from GRBs, and the cosmic ray-neutrino connection,''
  Astrophys.\ J.\  {\bf 768}, 186 (2013)
  [arXiv:1301.6163 [astro-ph.HE]].
  %%CITATION = ARXIV:1301.6163;%%
  %2 citations counted in INSPIRE as of 23 Aug 2013

%\cite{Zhang:2012qy}
\bibitem{Zhang:2012qy} 
  B.~Zhang and P.~Kumar,
  %``Model-dependent high-energy neutrino flux from Gamma-Ray Bursts,''
  Phys.\ Rev.\ Lett.\  {\bf 110}, 121101 (2013)
  [arXiv:1210.0647 [astro-ph.HE]].
  %%CITATION = ARXIV:1210.0647;%%
  %8 citations counted in INSPIRE as of 26 Aug 2013

%\cite{Wolfenstein:1981kw}
\bibitem{Wolfenstein:1981kw} 
  L.~Wolfenstein,
  %``Different Varieties of Massive Dirac Neutrinos,''
  Nucl.\ Phys.\ B {\bf 186}, 147 (1981).
  %%CITATION = NUPHA,B186,147;%%
  %211 citations counted in INSPIRE as of 06 Sep 2013

%\cite{Petcov:1982ya}
\bibitem{Petcov:1982ya} 
  S.~T.~Petcov,
  %``On Pseudodirac Neutrinos, Neutrino Oscillations and Neutrinoless Double beta Decay,''
  Phys.\ Lett.\ B {\bf 110}, 245 (1982).
  %%CITATION = PHLTA,B110,245;%%
  %254 citations counted in INSPIRE as of 06 Sep 2013

%\cite{Bilenky:1983wt}
\bibitem{Bilenky:1983wt} 
  S.~M.~Bilenky and B.~Pontecorvo,
  %``Neutrino Oscillations With Large Oscillation Length In Spite Of Large (majorana) Neutrino Masses?,''
  Sov.\ J.\ Nucl.\ Phys.\  {\bf 38}, 248 (1983)
  [Lett.\ Nuovo Cim.\  {\bf 37}, 467 (1983)]
  [Yad.\ Fiz.\  {\bf 38}, 415 (1983)].
  %%CITATION = SJNCA,38,248;%%
  %26 citations counted in INSPIRE as of 06 Sep 2013

%\cite{Kobayashi:2000md}
\bibitem{Kobayashi:2000md} 
  M.~Kobayashi and C.~S.~Lim,
  %``Pseudo Dirac scenario for neutrino oscillations,''
  Phys.\ Rev.\ D {\bf 64}, 013003 (2001)
  [hep-ph/0012266].
  %%CITATION = HEP-PH/0012266;%%
  %28 citations counted in INSPIRE as of 06 Sep 2013

%\cite{Cahill:1999yk}
\bibitem{Cahill:1999yk} 
  K.~E.~Cahill,
  %``Neutrinos are nearly Dirac fermions,''
  hep-ph/9912416.
  %%CITATION = HEP-PH/9912416;%%
  %10 citations counted in INSPIRE as of 02 Nov 2013

%\cite{Beacom:2003eu}
\bibitem{Beacom:2003eu}
  J.~F.~Beacom, N.~F.~Bell, D.~Hooper, J.~G.~Learned, S.~Pakvasa and T.~J.~Weiler,
  %``PseudoDirac neutrinos: A Challenge for neutrino telescopes,''
  Phys.\ Rev.\ Lett.\  {\bf 92}, 011101 (2004)
  [hep-ph/0307151].
  %%CITATION = HEP-PH/0307151;%%
  %71 citations counted in INSPIRE as of 26 Feb 2013

%\cite{Esmaili:2012ac}
\bibitem{Esmaili:2012ac}
  A.~Esmaili and Y.~Farzan,
  %``Implications of the Pseudo-Dirac Scenario for Ultra High Energy Neutrinos from GRBs,''
  JCAP {\bf 1212}, 014 (2012)
  [arXiv:1208.6012 [hep-ph]].
  %%CITATION = ARXIV:1208.6012;%%
  %3 citations counted in INSPIRE as of 25 Feb 2013

%\cite{Pakvasa:2012db}
\bibitem{Pakvasa:2012db} 
  S.~Pakvasa, A.~Joshipura and S.~Mohanty,
  %``Explanation for the low flux of high energy astrophysical
  %muon-neutrinos,''
Phys.\ Rev.\ Lett.\  {\bf 110}, 171802 (2013)
  [arXiv:1209.5630 [hep-ph]].
  %%CITATION = ARXIV:1209.5630;%%
  %6 citations counted in INSPIRE as of 23 Aug 2013


%\cite{Joshipura:2013yba}
\bibitem{Joshipura:2013yba} 
  A.~S.~Joshipura, S.~Mohanty and S.~Pakvasa,
  %``Pseudo-Dirac neutrinos via mirror-world and depletion of UHE neutrinos,''
  arXiv:1307.5712 [hep-ph].
  %%CITATION = ARXIV:1307.5712;%%
  %1 citations counted in INSPIRE as of 23 Aug 2013

%\cite{Esmaili:2012nz}
\bibitem{Esmaili:2012nz} 
  A.~Esmaili, F.~Halzen and O.~L.~G.~Peres,
  %``Constraining Sterile Neutrinos with AMANDA and IceCube Atmospheric Neutrino Data,''
  JCAP {\bf 1211}, 041 (2012)
  [arXiv:1206.6903 [hep-ph]].
  %%CITATION = ARXIV:1206.6903;%%
  %19 citations counted in INSPIRE as of 30 Oct 2013

%\cite{Komatsu:2010fb}
\bibitem{Komatsu:2010fb}
  E.~Komatsu {\it et al.}  [WMAP Collaboration],
  %``Seven-Year Wilkinson Microwave Anisotropy Probe (WMAP) Observations: Cosmological Interpretation,''
  Astrophys.\ J.\ Suppl.\  {\bf 192}, 18 (2011)
  [arXiv:1001.4538 [astro-ph.CO]].
  %%CITATION = ARXIV:1001.4538;%%
  %3531 citations counted in INSPIRE as of 26 Feb 2013

%\cite{ArkaniHamed:2005yv}
\bibitem{ArkaniHamed:2005yv} 
  N.~Arkani-Hamed, S.~Dimopoulos and S.~Kachru,
  %``Predictive landscapes and new physics at a TeV,''
  hep-th/0501082.
  %%CITATION = HEP-TH/0501082;%%
  %129 citations counted in INSPIRE as of 23 Aug 2013

%\cite{Mahbubani:2005pt}
\bibitem{Mahbubani:2005pt} 
  R.~Mahbubani and L.~Senatore,
  %``The Minimal model for dark matter and unification,''
  Phys.\ Rev.\ D {\bf 73}, 043510 (2006)
  [hep-ph/0510064].
  %%CITATION = HEP-PH/0510064;%%
  %36 citations counted in INSPIRE as of 23 Aug 2013

%\cite{D'Eramo:2007ga}
\bibitem{D'Eramo:2007ga} 
  F.~D'Eramo,
  %``Dark matter and Higgs boson physics,''
  Phys.\ Rev.\ D {\bf 76}, 083522 (2007)
  [arXiv:0705.4493 [hep-ph]].
  %%CITATION = ARXIV:0705.4493;%%
  %26 citations counted in INSPIRE as of 23 Aug 2013

%\cite{Enberg:2007rp}
\bibitem{Enberg:2007rp} 
  R.~Enberg, P.~J.~Fox, L.~J.~Hall, A.~Y.~Papaioannou and M.~Papucci,
  %``LHC and dark matter signals of improved naturalness,''
  JHEP {\bf 0711}, 014 (2007)
  [arXiv:0706.0918 [hep-ph]].
  %%CITATION = ARXIV:0706.0918;%%
  %16 citations counted in INSPIRE as of 23 Aug 2013

%\cite{Giunti:2011gz}
\bibitem{Giunti:2011gz}
  C.~Giunti and M.~Laveder,
  %``3+1 and 3+2 Sterile Neutrino Fits,''
  Phys.\ Rev.\ D {\bf 84}, 073008 (2011)
  [arXiv:1107.1452 [hep-ph]].
  %%CITATION = ARXIV:1107.1452;%%
  %66 citations counted in INSPIRE as of 26 Feb 2013

%\cite{Dodelson:1993je}
\bibitem{Dodelson:1993je} 
  S.~Dodelson and L.~M.~Widrow,
  %``Sterile-neutrinos as dark matter,''
  Phys.\ Rev.\ Lett.\  {\bf 72}, 17 (1994)
  [hep-ph/9303287].
  %%CITATION = HEP-PH/9303287;%%
  %379 citations counted in INSPIRE as of 31 Oct 2013

%\cite{Kusenko:1997sp}
\bibitem{Kusenko:1997sp} 
  A.~Kusenko and G.~Segre,
  %``Neutral current induced neutrino oscillations in a supernova,''
  Phys.\ Lett.\ B {\bf 396}, 197 (1997)
  [hep-ph/9701311].
  %%CITATION = HEP-PH/9701311;%%
  %138 citations counted in INSPIRE as of 31 Oct 2013

%\cite{Archidiacono:2012ri}
\bibitem{Archidiacono:2012ri}
  M.~Archidiacono, N.~Fornengo, C.~Giunti and A.~Melchiorri,
  %``Testing 3+1 and 3+2 neutrino mass models with cosmology and short baseline experiments,''
  Phys.\ Rev.\ D {\bf 86}, 065028 (2012)
  [arXiv:1207.6515 [astro-ph.CO]].
  %%CITATION = ARXIV:1207.6515;%%
  %15 citations counted in INSPIRE as of 26 Feb 2013


%\cite{Donini:2012tt}
\bibitem{Donini:2012tt}
  A.~Donini, P.~Hernandez, J.~Lopez-Pavon, M.~Maltoni and T.~Schwetz,
  %``The minimal 3+2 neutrino model versus oscillation anomalies,''
  JHEP {\bf 1207}, 161 (2012)
  [arXiv:1205.5230 [hep-ph]].
  %%CITATION = ARXIV:1205.5230;%%
  %11 citations counted in INSPIRE as of 26 Feb 2013

%\cite{Keranen:2003xd}
\bibitem{Keranen:2003xd}
  P.~Keranen, J.~Maalampi, M.~Myyrylainen and J.~Riittinen,
  %``Effects of sterile neutrinos on the ultrahigh-energy cosmic neutrino flux,''
  Phys.\ Lett.\ B {\bf 574}, 162 (2003)
  [hep-ph/0307041].
  %%CITATION = HEP-PH/0307041;%%
  %42 citations counted in INSPIRE as of 25 Feb 2013

%\cite{Biermann:2006bu}
\bibitem{Biermann:2006bu}
  P.~L.~Biermann and A.~Kusenko,
  %``Relic keV sterile neutrinos and reionization,''
  Phys.\ Rev.\ Lett.\  {\bf 96}, 091301 (2006)
  [astro-ph/0601004].
  %%CITATION = ASTRO-PH/0601004;%%
  %127 citations counted in INSPIRE as of 26 Feb 2013

%\cite{Conrad:2012qt}
\bibitem{Conrad:2012qt}
  J.~M.~Conrad, C.~M.~Ignarra, G.~Karagiorgi, M.~H.~Shaevitz and J.~Spitz,
  %``Sterile Neutrino Fits to Short Baseline Neutrino Oscillation Measurements,''
  arXiv:1207.4765 [hep-ex].
  %%CITATION = ARXIV:1207.4765;%%
  %13 citations counted in INSPIRE as of 26 Feb 2013

%\cite{Dev:2012bd}
\bibitem{Dev:2012bd}
  P.~S.~B.~Dev and A.~Pilaftsis,
  %``Light and Superlight Sterile Neutrinos in the Minimal Radiative
  %Inverse Seesaw Model,''
Phys.\ Rev.\ D {\bf 87}, 053007 (2013)
  arXiv:1212.3808 [hep-ph].
  %%CITATION = ARXIV:1212.3808;%%
  %2 citations counted in INSPIRE as of 26 Feb 2013

%\cite{Merle:2013gea}
\bibitem{Merle:2013gea} 
  A.~Merle,
  %``keV Neutrino Model Building,''
  Int.\ J.\ Mod.\ Phys.\ D {\bf 22}, 1330020 (2013)
  [arXiv:1302.2625 [hep-ph]].
  %%CITATION = ARXIV:1302.2625;%%
  %13 citations counted in INSPIRE as of 31 Oct 2013


%\cite{Gupta:2013vva}
\bibitem{Gupta:2013vva} 
  V.~Gupta, G.~Sánchez-Colón, S.~Rajpoot and H-C.~Wang,
  %``Lepton flavor mixing in the Wolfenstein scheme,''
  Phys.\ Rev.\ D {\bf 87}, 073009 (2013)
  [arXiv:1304.1065 [hep-ph]].
  %%CITATION = ARXIV:1304.1065;%%


%\cite{Kraus:2004zw}
\bibitem{Kraus:2004zw}
  C.~Kraus, B.~Bornschein, L.~Bornschein, J.~Bonn, B.~Flatt, A.~Kovalik, B.~Ostrick and E.~W.~Otten {\it et al.},
  %``Final results from phase II of the Mainz neutrino mass search in tritium beta decay,''
  Eur.\ Phys.\ J.\ C {\bf 40}, 447 (2005)
  [hep-ex/0412056].
  %%CITATION = HEP-EX/0412056;%%
  %286 citations counted in INSPIRE as of 26 Feb 2013

%\cite{Arnold:2005rz}
\bibitem{Arnold:2005rz}
  R.~Arnold {\it et al.}  [NEMO Collaboration],
  %``First results of the search of neutrinoless double beta decay with the NEMO 3 detector,''
  Phys.\ Rev.\ Lett.\  {\bf 95}, 182302 (2005)
  [hep-ex/0507083].
  %%CITATION = HEP-EX/0507083;%%
  %133 citations counted in INSPIRE as of 26 Feb 2013

%\cite{Bongrand:2011ei}
\bibitem{Bongrand:2011ei}
  M.~Bongrand [NEMO-3 Collaboration],
  %``Results of the NEMO-3 Double Beta Decay Experiment,''
  arXiv:1105.2435 [hep-ex].
  %%CITATION = ARXIV:1105.2435;%%
  %3 citations counted in INSPIRE as of 26 Feb 2013


%\cite{Fox:2006iu}
\bibitem{Fox:2006iu}
  D.~B.~Fox and P.~Meszaros,
  %``GRB Fireball Physics: Prompt and Early Emission,''
  New J.\ Phys.\  {\bf 8}, 199 (2006)
  [astro-ph/0609173].
  %%CITATION = ASTRO-PH/0609173;%%
  %11 citations counted in INSPIRE as of 26 Feb 2013

%\cite{Learned:1994wg}
\bibitem{Learned:1994wg} 
  J.~G.~Learned and S.~Pakvasa,
  %``Detecting tau-neutrino oscillations at PeV energies,''
  Astropart.\ Phys.\  {\bf 3}, 267 (1995)
  [hep-ph/9405296, hep-ph/9408296].
  %%CITATION = HEP-PH/9405296,;%%
  %331 citations counted in INSPIRE as of 30 Oct 2013


%\cite{Rachen:1998fd}
\bibitem{Rachen:1998fd} 
  J.~P.~Rachen and P.~Meszaros,
  %``Photohadronic neutrinos from transients in astrophysical sources,''
  Phys.\ Rev.\ D {\bf 58}, 123005 (1998)
  [astro-ph/9802280].
  %%CITATION = ASTRO-PH/9802280;%%
  %205 citations counted in INSPIRE as of 30 Oct 2013

%\cite{Lipari:2007su}
\bibitem{Lipari:2007su} 
  P.~Lipari, M.~Lusignoli and D.~Meloni,
  %``Flavor Composition and Energy Spectrum of Astrophysical Neutrinos,''
  Phys.\ Rev.\ D {\bf 75}, 123005 (2007)
  [arXiv:0704.0718 [astro-ph]].
  %%CITATION = ARXIV:0704.0718;%%
  %65 citations counted in INSPIRE as of 30 Oct 2013

%\cite{Pakvasa:2007dc}
\bibitem{Pakvasa:2007dc} 
  S.~Pakvasa, W.~Rodejohann and T.~J.~Weiler,
  %``Flavor Ratios of Astrophysical Neutrinos: Implications for Precision Measurements,''
  JHEP {\bf 0802}, 005 (2008)
  [arXiv:0711.4517 [hep-ph]].
  %%CITATION = ARXIV:0711.4517;%%
  %39 citations counted in INSPIRE as of 30 Oct 2013

%\cite{Kashti:2005qa}
\bibitem{Kashti:2005qa}
  T.~Kashti and E.~Waxman,
  %``Flavoring astrophysical neutrinos: Flavor ratios depend on energy,''
  Phys.\ Rev.\ Lett.\  {\bf 95}, 181101 (2005)
  [astro-ph/0507599].
  %%CITATION = ASTRO-PH/0507599;%%
  %91 citations counted in INSPIRE as of 26 Feb 2013

%\cite{Crocker:2004bb}
\bibitem{Crocker:2004bb} 
  R.~M.~Crocker, M.~Fatuzzo, R.~Jokipii, F.~Melia and R.~R.~Volkas,
  %``The AGASA/SUGAR anisotropies and TeV gamma rays from the Galactic Center: A Possible signature of extremely high-energy neutrons,''
  Astrophys.\ J.\  {\bf 622}, 892 (2005)
  [astro-ph/0408183].
  %%CITATION = ASTRO-PH/0408183;%%
  %63 citations counted in INSPIRE as of 31 Oct 2013

%\cite{Halzen:2011yq}
\bibitem{Halzen:2011yq}
  F.~Halzen,
  %``Sterile Neutrinos and IceCube,''
  J.\ Phys.\ Conf.\ Ser.\  {\bf 408}, 012023 (2013)
  [arXiv:1111.0918 [hep-ph]].
  %%CITATION = ARXIV:1111.0918;%%
  %5 citations counted in INSPIRE as of 26 Feb 2013





\end{thebibliography}
\end{document}